%% file: 00_arxiv_threshold_result_for_2wcsts.tex
\documentclass[11pt]{article}

\input{packages}
\input{macros_letters}

\input{macros}

\graphicspath{{./}}

\usepackage{array}
\usepackage[title]{appendix}
\usepackage{graphicx}
\usepackage{tikz}
\usetikzlibrary{automata,positioning} \usepackage{dsfont} \usepackage{xspace}
\usepackage[nothing]{algorithm}
\usepackage{algorithmicx}
\usepackage[noend]{algpseudocode}
\algblockdefx[protocol]{Protocol}{EndProtocol}{\textbf{Protocol}\ }{}

\floatname{algorithm}{Pseudocode}

\title{A Tight Threshold Bound for Search Trees with 2-way Comparisons\thanks{Research supported by NSF grant CCF-2153723.}}

\author[$\dagger$]{Sunny Atalig}
\author[$\dagger$]{Marek Chrobak}

\affil[$\dagger$]{University of California at Riverside, USA}

\begin{document}

\maketitle

\begin{abstract}
We study search trees with 2-way comparisons (\twoWCST's), which involve
separate less-than and equal-to tests in their nodes, each test having two possible outcomes, yes and no.
These trees have a much subtler structure than standard
search trees with 3-way comparisons (\threeWCST's) and are still not well understood,
hampering progress towards designing an efficient algorithm for computing minimum-cost trees.
One question that attracted attention in the past is whether there is
an easy way to determine which type of comparison should be applied at any step of the search.
Anderson, Kannan, Karloff and Ladner studied this in terms of the ratio between the
maximum and total key weight, and defined two threshold values:
$\lowthreshold$ is the largest ratio that forces the less-than test,
and $\highthreshold$ is the smallest ratio that forces the equal-to test. They determined that
$\lowthreshold = \onefourth$, but for the higher threshold they only showed that 
$\highthreshold\in [\threesevenths,\fournineths]$.
We give the tight bound for the higher threshold, by proving that in fact
$\highthreshold = \threesevenths$.
\end{abstract}


\section{Introduction}
\label{sec: introduction}

\input{01_introduction.tex}


\section{Preliminaries}
\label{sec: preliminaries}

\input{02_preliminaries.tex}


\section{Proof of Theorem~\ref{theorem: lambda+ upper bound}}
\label{sec: Proof of threshold bound}

\input{03_proof_of_threshold_bound.tex}


\bibliographystyle{plain}
\bibliography{search_trees}

\end{document}

%% file: packages.tex
\usepackage[margin=2.5cm]{geometry}
\usepackage{authblk}

\usepackage{amssymb,amsmath,amsthm,amsfonts,thmtools}
\usepackage{color}

\usepackage{yfonts}
\usepackage{mathrsfs}

\usepackage{etoolbox} 

\usepackage{graphicx}
\usepackage[export]{adjustbox}
\usepackage{xspace}

\usepackage{times}
\usepackage{verbatim}
\usepackage{enumitem}
\usepackage{graphicx,wrapfig,lipsum}
\usepackage[font=small]{caption}
\usepackage{subcaption}
\usepackage[english]{babel}
\usepackage[utf8]{inputenc}

\usepackage[noend]{algpseudocode}
\usepackage{url}
\usepackage[all]{xy}
\usepackage{rotating}
\usepackage{ifpdf}
\usepackage{tcolorbox}
\usepackage{lipsum}

\usepackage{mdframed}             
\usepackage{xifthen} 
\usepackage{mathtools}  

\DeclareMathAlphabet\mathbfcal{OMS}{cmsy}{b}{n}

\usepackage{comment}
\usepackage[T1]{fontenc}

\usepackage[hidelinks]{hyperref} 

%% file: macros_letters.tex
\newcommand{\barA}{{\bar A}}

\newcommand{\calK}{\ensuremath{\mathcal K}\xspace}



%% file: macros.tex
\colorlet{darkgreen}{green!45!black}

\newtheorem{theorem}{Theorem}[section]
\newtheorem{lemma}{Lemma}[section]

\newtheorem{claim}[lemma]{Claim}

\newcounter{case}
\newcounter{subcase}[case]
\newcounter{subsubcase}[subcase]
\newenvironment{case}[1][\unskip]{
\paragraph{Case \thecase:} #1.
}

\newenvironment{subcase}[1][\unskip]{
\paragraph{Case \thecase.\thesubcase:} #1.
}

\AtBeginEnvironment{case}{\refstepcounter{case}}
\AtBeginEnvironment{proof}{\setcounter{case}{0}}
\AtBeginEnvironment{subcase}{\refstepcounter{subcase}}
\AtBeginEnvironment{subsubcase}{\refstepcounter{subsubcase}}

\makeatletter
\renewcommand{\p@subcase}{\thecase.}
\renewcommand{\p@subsubcase}{\thecase.\thesubcase}
\makeatother

\newcommand{\ignore}[1]{}

\newcommand{\margincomment}[2]%
{\marginpar{\footnotesize\raggedright {\color{red}#1}: #2}}

\newcommand{\etal}{et~al.\ }

\newcommand{\braced}[1]{{ \left\{ {#1} \right\} }}

\newcommand{\angled}[1]{{ \langle {#1} \rangle }}

\newcommand{\parend}[1]{{ \left({#1} \right) }}

\newcommand{\EqOrGe}{=\!\!/\!\!\ge}
\newcommand{\NeOrLe}{\ne\!\!/\!\!<}
\newcommand{\Keys}{\calK}
\newcommand{\sideweight}{\textit{sw}}

\newcommand{\threeWCST}{\textsc{3wcst}\xspace}
\newcommand{\twoWCST}{\textsc{2wcst}\xspace}

\DeclareMathOperator{\cost}{cost}
\DeclareMathOperator{\depth}{depth}
\newcommand{\highlambda}{\lambda^+}
\newcommand{\lowlambda}{\lambda^-}

\newcommand{\lowthreshold}{\lowlambda}
\newcommand{\highthreshold}{\highlambda}
\newcommand{\equalstest}[1]{\angled{=#1}}
\newcommand{\lessthantest}[1]{\angled{<#1}}
\newcommand{\unknowntest}[1]{\angled{\mathrel{\ast}#1}}
\newcommand{\containtest}[1]{\angled{\in #1}}

\newcommand{\onethird}{\tfrac{1}{3}}
\newcommand{\seventhirds}{\tfrac{7}{3}}

\newcommand{\onefourth}{\tfrac{1}{4}}

\newcommand{\threesevenths}{\tfrac{3}{7}}
\newcommand{\oneseventh}{\tfrac{1}{7}}
\newcommand{\fournineths}{\tfrac{4}{9}}

\newcommand{\ListLengths}{\setlength{\itemsep}{0ex}\setlength{\topsep}{1ex}\setlength{\partopsep}{0ex}}



%% file: 01_introduction.tex



Search trees are decision-tree based data structures used for identifying a query value within some
specified set $\Keys$ of keys, by applying some simple tests on the query.
When $\Keys$ is a linearly ordered set, these tests can be comparisons between the query and a key from $\Keys$.
In the classical model of 3-way comparison trees (\threeWCST's), each comparison has three outcomes:
the query can be smaller, equal, or greater than the key associated with this node. 
In the less studied model of search trees with 2-way comparisons (\twoWCST's), proposed by
Knuth~\cite[\S 6.2.2, Example 33]{Knuth1998},
we have separate less-than and equal-to tests, with each test having two outcomes, yes or no. In both models,
the search starts at the root node of the tree and proceeds down the tree, following the branches 
corresponding to these outcomes, until the query ends up in the leaf representing the key
equal to the query value~\footnote{Here we assume the scenario when the query is in $\Keys$,
often called the \emph{successful-query} model. If arbitrary queries are allowed, the tree
also needs to have leaves representing inter-key intervals. Algorithms for the
successful-query model typically extend naturally to this general mode without increasing
their running time.}.

The focus of this paper is on the static scenario, when the key set $\Keys$ does not change over time\footnote{%
There is of course vast amount of research on the dynamic case, when the goal is to have the tree to adapt
to the input sequence, but it is not relevant to this paper.}.
Each key $k$ is assigned a non-negative weight $w_k$, representing the frequency of $k$, or the 
probability of it appearing on input.
Given these weights, the objective is to compute a tree $T$ that minimizes its cost,
defined by $\cost(T) = \sum_{k\in\Keys} w_k\cdot\depth(k)$, where $\depth(k)$ is the depth of the 
leaf representing key $k$ in $T$. This concept naturally captures the expected cost of a random query.

\begin{figure}[t]
	\centering
	\includegraphics[width=2.5in]{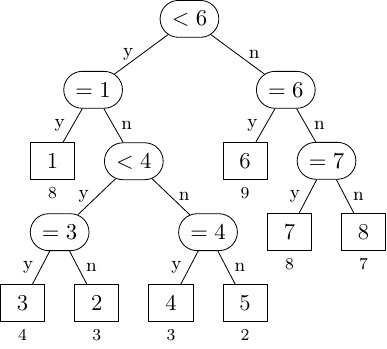}
	\caption{ An example of a \twoWCST. This tree handles keys 1, 2, 3, 4, 5, 6, 7, 8 with respective weights 8, 3, 4, 3, 2, 9, 8, 7. Computing the cost in terms of leaves, we get $8 \cdot 2 + 3 \cdot 4 + 4 \cdot 4 + 3\cdot 4 + 2 \cdot 4 + 9 \cdot 2 + 8\cdot 3 + 7\cdot 3 = 127$. 
	} 
	\label{fig: tree-example}
\end{figure}

Optimal \threeWCST's can be computed in time $O(n^3)$ using the now standard dynamic programming
algorithm. As shown already by Knuth~\cite{Knuth1971} in 1971
and later by Yao~\cite{yao_efficient_dynamic_programming_80} in 1980, using a more general approach,
the running time can be improved to $O(n^2)$. This improvement
leverages the property of \threeWCST's called the quadrangle inequality (which is
essentially equivalent to the so-called Monge property or total monotonicity --- see~\cite{bein_et_al_knuth-yao_total_monotonicity_09}).
In contrast, the first (correct\footnote{%
Anderson~\etal~\cite{Anderson2002} reference an earlier $O(n^5)$-time algorithm
by Spuler~\cite{Spuler1994Paper,Spuler1994Thesis}. However, as shown in~\cite{chrobak_huang_2022}, Spuler's proof
is not correct.
}) 
polynomial-time algorithm for
finding optimal \twoWCST's was developed only in 2002 by Anderson, Kannan, Karloff and Ladner~\cite{Anderson2002}.
Its running time is $O(n^4)$. A simpler and slightly more general 
$O(n^4)$-time algorithm was recently given in~\cite{chrobak_simple_2021}.

The reason for this disparity in the running times lies in the internal structure of
\twoWCST's, which is much more intricate than that of \threeWCST's. Roughly, while
the optimal cost function of \threeWCST's has a dynamic-programming formulation where all sub-problems
are represented by key intervals, this is not the case for \twoWCST's.
For \twoWCST's, a similar approach produces intervals with holes (corresponding to earlier failed
equal-to tests), leading to exponentially many sub-problems.
As shown in~\cite{Anderson2002} (and conjectured earlier by Spuler~\cite{Spuler1994Paper,Spuler1994Thesis}), 
this can be reduced to $O(n^3)$ sub-problems using the so-called heaviest-first key property.

One other challenge in designing a faster algorithm is that, for any sub-problem, it is not known
a priori whether the root should use the less-than test or the equal-to test.
The intuition is that when some key is sufficiently heavy then the optimum tree must
start with the equal-to test (to this key) at the root.
On the other hand, if all weights are roughly equal (and there are sufficiently many keys), then the tree should start with
a less-than test, to break the instance into two parts with roughly the same total weight.
Addressing this, Anderson et al.~\cite{Anderson2002}
introduced two threshold values for the maximum key weight. Denoting by $W$ the total
weight of the keys in the instance, these values are defined as follows:
\begin{itemize}\setlength{\itemsep}{0.025in}
	\item $\lowthreshold$ is the largest $\lambda$ such that if all key weights are smaller than
			$\lambda W$ then there is no optimal tree with an equal-to test at the root.
	\item $\highthreshold$ is the smallest $\lambda$ such that if any key
			has weight at least $\lambda W$ then there is an optimal tree with an equal-to test at the root.
\end{itemize}
In their paper\footnote{In~\cite{Anderson2002} the notation for the threshold values $\lowthreshold$ and $\highthreshold$ was, respectively, $\lambda$ and $\mu$. 
We changed the notation to make it more intuitive and consistent with~\cite{atalig_eal_structural_properties_23}.}, 
they proved that $\lowthreshold = \onefourth$ and $\highthreshold \in [\threesevenths,\fournineths]$.
These thresholds played a role in their $O(n^4)$-time algorithm for computing optimal \twoWCST's.
The more recent $O(n^4)$-time algorithm in~\cite{chrobak_simple_2021} uses a
somewhat different approach and does not rely on any threshold bounds on key weights.

Nevertheless, breaking the $O(n^4)$ barrier will require deeper understanding of the
structure of optimal \twoWCST's; in particular, more accurate criteria for 
determining which of the two tests should be applied first are 
likely to  be useful. Even if not improving the asymptotic complexity, 
such criteria reduce computational overhead by limiting the number of keys to be
considered for less-than tests. 

With these motivations in mind, in this work we give a tight bound on the higher weight threshold,
by proving that the lower bound of $\threesevenths$ for $\highthreshold$ in~\cite{Anderson2002} is in fact tight.


\begin{theorem} \label{theorem: lambda+ upper bound}
For all $n \ge 2$, if an instance of $n$ keys has total weight $W$ and
a maximum key weight at least $\threesevenths W$
then there exists an optimal \twoWCST rooted at an equal-to test. In other words, $\highlambda  \le \threesevenths $.
\end{theorem}

The proof is given in Section~\ref{sec: Proof of threshold bound}, after we introduce the
necessary definitions and notation, and review fundamental properties of \twoWCST's, in Section~\ref{sec: preliminaries}.

\smallskip

\emph{Note:}
For interested readers, other structural properties of \twoWCST's were recently studied in the companion paper~\cite{atalig_eal_structural_properties_23}.
In particular, that paper provides other types of threshold bounds, including one that involves two heaviest keys,
as well as examples showing that the speed-up techniques for dynamic programming, including the quadrangle inequality,
do not work for \twoWCST's.

%% file: 02_preliminaries.tex



\paragraph{Notation.} 
Without any loss of generality we can assume that set of keys is $\Keys = \braced{1,2,...,n}$,
and their corresponding weights are denoted $w_1 , w_2, \ldots, w_n$.
Throughout the paper, we will typically use letters $i,j,k,...$ to represent keys. 
The total weight of the instance is denoted  by $W = \sum_{k\in\Keys} w_k$.  For a tree  $T$ by
$w\parend{T}$ we denote the total weight of keys in its leaves, calling it the \emph{weight of $T$}.
For a node $v$ in $T$, $w\parend{v}$ denotes the weight of the sub-tree of $T$ rooted at $v$.

Each internal node is either an equal-to test to a key $k$, denoted by $\equalstest{k}$,
or a less-than test to $k$, denoted by  $\lessthantest{k}$. 
Conventionally, the left and right branches of the tree at a node are labelled by ``yes'' and ``no'' outcomes,
but in our proof we will often depart from this notation and use relation symbols
``$=$'', ``$\neq$'', ``$<$'', etc, instead. 
For some nodes only the comparison key $k$ will be specified, but not the test type.
Such nodes will be denoted $\unknowntest{k}$, and their
outcome branches will be labeled ``$\EqOrGe$'' and ``$\NeOrLe$''.
The interpretation of these is natural: If $\unknowntest{k}$ is $\equalstest{k}$
then the first branch is taken on they ``yes'' answer, otherwise the second branch is taken.
If $\unknowntest{k}$ is $\lessthantest{k}$ then the second branch is taken on the ``yes'' answer
and otherwise the first branch is taken.
This convention will be very useful in reducing the case complexity.

A branch of a node is called \emph{redundant} if there is no query value $q\in \Keys$
that traverses this branch during search. A \twoWCST $T$ is called \emph{irreducible}
if it does not contain any redundant branches. Each tree can be converted into an irreducible
tree by ``splicing out'' redundant branches (linking the sibling branch drectly to the grandparent).
This does not increase cost. So throughout the paper we tacitly assume that any given tree is irreducible.
In particular, note that any key $k$ appearing in a node $\unknowntest{k}$ of an irreducible tree must satisfy
all outcomes of the tests on the path from the root to $\unknowntest{k}$.
(The only non-trivial case is when $\unknowntest{k}$ is $\lessthantest{k}$ and there is a node $\equalstest{k}$
along this path. In this case, $\lessthantest{k}$ can be replaced by $\lessthantest{k+1}$, as in this case
$k$ cannot be $n$ if $T$ is irreducible.)


\paragraph{Side weights.} 
We use some concepts and auxiliary lemmas developed by Anderson et al.~\cite{Anderson2002}.
In particular, the concept of side-weights is useful.
The \emph{side-weight} of a node $v$ in a \twoWCST $T$ is defined by
	\[ \sideweight\parend{v} = \begin{cases}
		0  &  \text{if } v \text{ is a leaf} \\
		w_k &  \text{if } v \text{ is an equal-to test $\equalstest{k}$ } \\
		\min\braced{w\parend{L} , w\parend{R}} & \text{if } v \text{ is a less-than test with sub-trees } L, R 
	\end{cases}\]


\begin{lemma} \label{lemma: side-weight monoticity}
	\emph{\cite{Anderson2002}}
	Let \(T\) be an optimal \twoWCST. Then \(\sideweight\parend{u} \ge \sideweight\parend{v}\) if \(u\) is a parent of \(v\).
\end{lemma}

The lemma, while far from obvious, can be proved by applying so-called ``rotations'' to
the tree, which are local rearrangements that swap some adjacent nodes.
As a simple example, if $u = \equalstest{k}$ is a child of $v = \equalstest{l}$ and
$sw(u) > sw(v)$ then exchanging these two comparisons would reduce the tree cost, contradicting optimality.
For the complete proof, see~\cite{Anderson2002}. 

\smallskip

Lemma~\ref{lemma: side-weight monoticity} implies immediately the following:


\begin{lemma} \label{lemma: RMLK property}
	\emph{\cite{Anderson2002}}
	For \(n > 2\), if an optimal \twoWCST for an instance of \(n\) keys is rooted at an equal-to test on \(i\), 
	then \(i\) is a key of maximum-weight.
\end{lemma}

We remark that in case of ties between maximum-weight keys a subtlety arises that led to
some complications in the algorithm in~\cite{Anderson2002}. It was shown later in~\cite{chrobak_simple_2021}
that this issue can be circumvented. This issue does not arise in our paper, and the above statement of
Lemma~\ref{lemma: RMLK property} is sufficient for our argument.


\paragraph{More about rotations.} 
We will extend the concept of rotations to involve nodes of the form $\unknowntest{k}$, with unspecified tests.
This makes the rotations somewhat non-trivial, since such nodes represent two different possible subtrees, and
we need to justify that the tree obtained from the rotation is correct in both cases. 

\begin{figure}[t]
	\centering
	\includegraphics[valign=m,width=1.25in]{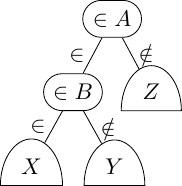}\qquad$\boldsymbol{\Longrightarrow}$\qquad
	\includegraphics[valign=m,width=1.25in]{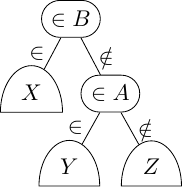}
	\caption{A tree rotation using general queries.
	} 
	\label{fig: generic query}
\end{figure}

\begin{figure}[t]
	\centering
	\begin{subfigure}{0.48\textwidth}
		\centering
		\includegraphics[valign=m,width=1.25in]{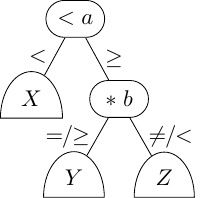}\quad$\boldsymbol{\Longrightarrow}$\quad
		\includegraphics[valign=m,width=1.25in]{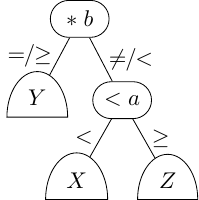}
		\caption{Valid if tree is irreducible.} 
				\label{fig: double outcome 1}
	\end{subfigure}
	\begin{subfigure}{0.48\textwidth}
		\centering
		\includegraphics[valign=m,width=1.25in]{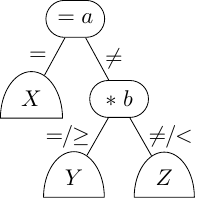}\quad$\boldsymbol{\Longrightarrow}$\quad
		\includegraphics[valign=m,width=1.25in]{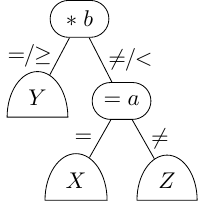}
		\caption{Valid if $a<b$ or $\unknowntest{b} = \equalstest{b}$.} 
		\label{fig: double outcome 2}
	\end{subfigure}
	\caption{Rotating a node with double outcomes.} 
	\label{fig: double outcome}
\end{figure}

To give a simple condition for ensuring valid rotations, we generalize \twoWCST{s} by considering arbitrary types of tests, not merely equal-to or less-than tests. 
Any binary test can be identified by the set of keys that satisfy the ``yes'' outcome. 
We can thus represent such a test using the set element relation, denoted $\containtest{A}$, with branches labelled ``$\in$'' and ``$\notin$''. 
An equal-to test $\equalstest{k}$ can be identified by the singleton set $\braced{k}$ or its complement ${\Keys} - \braced{k}$, 
and a less-than test $\lessthantest{k}$ can be identified by $\parend{-\infty, k}$ or $[k, \infty)$.

Consider the rotation in Figure~\ref{fig: generic query}, where we denote the sub-trees and the keys they contain by $X$, $Y$, and $Z$. 
Comparing the trees on the left and the right, we must have $X = A\cap B = B$, which implies that for the rotation to be
correct we need $B\subseteq A$. (If the tree on the left is irreducible, the containment must in fact be strict.)
On the other hand, if $B\subseteq A$ then in both trees we have $Y = A\setminus B$ and $Z = \Keys-A$.
This gives us the following property:

\begin{description}
\item{\textbf{Containment Property:}} The rotation shown in Figure~\ref{fig: generic query} is valid if and only if $B\subseteq A$.
\end{description}

We remark that other rotations, such as when $\containtest{B}$ is in the $\notin$-branch of $\containtest{A}$, can be accounted for by the 
above containment property, by replacing $\containtest{A}$ with $\containtest{\barA}$, where $\barA$ is the complement of $A$.

Consider Figure~\ref{fig: double outcome 1}, with tests $\lessthantest{a}$ and $\unknowntest{b}$. 
Assuming the original tree is irreducible, we have $b \ge a$, with strict inequality if $\unknowntest{b}$ is a less-than test. 
We identify $\lessthantest{a}$ with set $A = [a, \infty)$ and $\unknowntest{b}$ with $B = \braced{b}$ or $[b, \infty)$ (second option corresponds to $\lessthantest{b}$). 
Then by our inequalities, it is clear that $B \subset A$ and the rotation is valid. By a similar reasoning, the rotation shown in Figure~\ref{fig: double outcome 2} is also valid, assuming either $a<b$ or $\unknowntest{b}$ is an equal-to test. For proving Theorem~\ref{theorem: lambda+ upper bound}, we need only consider these two rotations (or the corresponding reverse rotations).

\smallskip

Not all tree modifications in our proof are rotations. Some modifications also
\emph{insert} a new comparison test into the tree. This modification has the effect of making a tree reducible, 
though it can be converted into a irreducible tree, as explained earlier in this section. 
Insertions are used by Anderson et al.~\cite{Anderson2002} in proving 
the tight bound for $\lowlambda$ and will also be used in proving Theorem~\ref{theorem: lambda+ upper bound}

%% file: 03_proof_of_threshold_bound.tex


In this section we prove Theorem~\ref{theorem: lambda+ upper bound}, namely 
that the lower bound of $\threesevenths$ on $\highlambda$ in~\cite{Anderson2002} is tight.

Before proceeding with the proof, we remind the reader that  $\unknowntest{k}$ denotes an unspecified comparison test on key $k$, 
and that its outcomes are specified with labels``$\EqOrGe$'' and ``$\NeOrLe$''.
	
	The proof is by induction on the number of nodes $n$. The cases $n = 2, 3$ are trivial so we'll move on to the inductive step. 
	Assume that $n \ge 4$ and that Theorem \ref{theorem: lambda+ upper bound} holds for all instances where the number of keys is in the range $\braced{2,\ldots, n-1}$.
	
	To show that the theorem holds for any $n$-key instance, consider  
	a tree $T$ for an instance with $n$ keys and whose maximum-weight key $m$ satisfies $w_m \ge \threesevenths w\parend{T}$, 
	and suppose that the root of $T$ is a 
	less-than test $\lessthantest{r}$. We show that we can then find another tree $T'$ that is rooted at an equal-to test node 
	and has cost not larger than $T$.
	
	If $r\in\braced{2,n}$, then one of the children of $\lessthantest{r}$ is a leaf, and we can simply replace $\lessthantest{r}$
	by an appropriate equal-to test.
	So we can assume that $2 < r \le n-1 $, in which case each sub-tree of $\lessthantest{r}$ must have between $2$ and $n-1$ nodes. 
	By symmetry, we also can assume that a heaviest-weight key $m$ is in the left sub-tree $L$ of $ \angled{< r }$.
	 Since $w_m \ge \threesevenths w\parend{T} \ge \threesevenths w\parend{L}$, we can replace $L$ with a tree
	 rooted at $\equalstest{m}$ without increasing cost, by our inductive assumption. 
	 (A careful reader might notice that, if a tie arises, the inductive assumption only guarantees that the 
	 root of this left subtree will be an equal-to test to a key of the same weight
	 as $m$. For simplicity, we assume that this key is $m$, for otherwise we can just use the other key in the rest of the proof.)
	 Let $T_1$ be the $\ne$-branch of $\equalstest{m}$ and $T_2$ be the $\ge$-branch of $\lessthantest{r}$.
	
\begin{figure}[t]
		\centering
		\begin{subfigure}[b]{0.48\textwidth}
			\centering
			\includegraphics[scale=0.75]{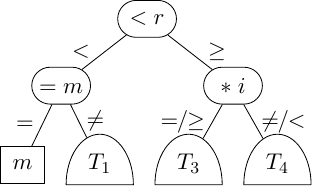}\caption{Original tree $T$}\label{fig: high lambda opt tree}
		\end{subfigure}
	\begin{subfigure}[b]{0.48\textwidth}
		\centering
		\includegraphics[scale=0.75]{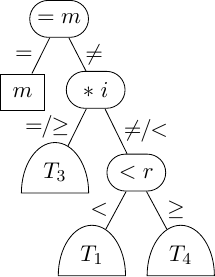}\caption{Modified tree $T'$}\label{fig: high lambda case 1}
	\end{subfigure}
		\caption{On the left, tree $T$ in the proof of Theorem~\ref{theorem: lambda+ upper bound}.
		On the right,
		the modification for Case~\ref{case 1: 3/7 bound}. Notice that $T_3$ is either the leaf $i$ or 
		contains all keys greater than or equal to $i$, 
		depending on the comparison test $\unknowntest{i}$.} \label{fig: high lambda fig 2}
	\end{figure}

	By applying Lemma~\ref{lemma: side-weight monoticity} to node $\equalstest{m}$ and its parent $\lessthantest{r}$,
	we have that $w\parend{T_2} \ge w_m \ge \threesevenths  w\parend{T}$, which in turn implies that $w\parend{T_1} \le \oneseventh w\parend{T}$. 
	Since $T_2$ is not a leaf, let its root be $\unknowntest{i}$  with sub-trees $T_3$ and $T_4$, 
	where $T_3$ is the $\EqOrGe$-branch. 
	The structure of $T$ is illustrated in Figure~\ref{fig: high lambda opt tree}.
	

	\smallskip
	
	To modify the tree, we break into two cases based on $T_3$'s weight:


	\begin{case}[$w\parend{T_3} \ge \onethird w\parend{T_2}$]\label{case 1: 3/7 bound}
		To obtain $T'$,
		rotate $\equalstest{m}$ to the root of the tree, then rotate $\unknowntest{i}$ so that it is the ${\ne}$-branch of $\angled{= m }$. 
		(The irreducibility of $T$ implies that the Containment Property in Section~\ref{sec: preliminaries} holds for both rotations.)
		Node $m$ goes up by $1$ in depth and subtrees $T_1$ and $T_4$ both go down by $1$ (see Figure~\ref{fig: high lambda case 1}). So the total change in cost is 
		
		\begin{align*}
			\cost(T')-\cost(T) \;&=\; w\parend{T_1} + w\parend{T_4} - w_m \\
			&=\; w\parend{T} - 2w_m - w\parend{T_3}   			&& w_m + w\parend{T_1} + w\parend{T_3} + w\parend{T_4} = w\parend{T} \\
			&\le\; w\parend{T} - 2w_m - \tfrac{1}{3} w_m  		&& w\parend{T_3} \ge \tfrac{1}{3} w\parend{T_2} , w\parend{T_2} \ge w_m  \\
			&=\; w\parend{T} - \seventhirds w_m \\
			&\le\; 0 				&& w_m \ge \threesevenths w\parend{T}  
		\end{align*}
	\end{case}
	
	\begin{case}[$w\parend{T_3} < \onethird w\parend{T_2}$]\label{case 2: 3/7 bound}
		Note that in this case, $w\parend{T_4} > \frac{2}{3}w\parend{T_2}$. We'll first handle some trivial cases. 
		If $T_4$ is a leaf $j$, we can replace $\unknowntest{i}$ with $\equalstest{j}$ and do the same rotations as in 
		Case~\ref{case 1: 3/7 bound} (swapping the roles of $T_4$ and $T_3$). 
		If $T_4$ contains only two leaves, say $k$ and $j$, applying Lemma~\ref{lemma: side-weight monoticity} we
		obtain that $w_k , w_j \le w\parend{T_3}$, which in turn implies that $w\parend{T_2} \le 3w\parend{T_3} < w\parend{ T_2}$ --- a contradiction.
		
		Therefore, we can assume that $T_4$ contains at least 3 leaves and at least 2 comparison nodes.
		 Let $\unknowntest{j}$ be the test in the root of $T_4$. 
		If $\unknowntest{j}$ is a less-than test and any of its branches is a leaf, we can replace $\unknowntest{j}$ with an equal-to test.
		If $\unknowntest{j}$ is an equal-to test than its  $\neq$-branch is not a leaf (because $T_4$ has at least 3 leaves).
		Thus we can assume that the $\NeOrLe$-branch of $\unknowntest{j}$ is not a leaf, and
		let $\unknowntest{k}$ be the root of this branch.
		This means that both $\unknowntest{j}$ and $\unknowntest{k}$  follow from an $\NeOrLe$-outcome. 
		Let  $T_5$ be the $\EqOrGe$-branch of $\unknowntest{j}$ and $T_6$ and $T_7$ be the branches of 	$\unknowntest{k}$. 
		The structure of $T$ is shown in Figure~\ref{fig: high lambda case 2.1}a.

		This idea behind the remainng argument is this: We now have that $T_5$, $T_6$ and $T_7$ together are relatively heavy (because $T_4$ is),
		while $T_1$ and $T_3$, which are at lower depth, are light. 
		If this weight difference is sufficiently large, it should be thus possible to rebalance the tree and reduce the cost.
		We will accomplish this rebalancing by using keys $i,j$ and $k$, although it may require introducing a new less-than test to one
		of these keys.
		
		For convenience, re-label the keys $\braced{i, j, k} = \braced{b_1 , b_2 , b_3}$ such that $b_1 \le b_2 \le b_3$. 
		The goal is to use $\lessthantest{b_2}$ as a ``central'' cut-point to divide the tree, with $\lessthantest{r}$ and $\unknowntest{b_1}$ in its $<$-branch.
		The  $\ge$-branch will contain $\unknowntest{b_3}$ and, if $\unknowntest{b_2}$ is an equal-to test, also $\unknowntest{b_2}$.


		\begin{figure}[t]
			\centering
			\includegraphics[valign=m,scale=0.75]{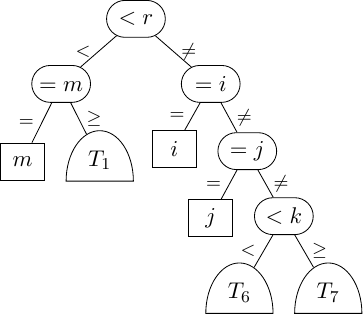}\qquad$\boldsymbol{\Longrightarrow}$\qquad
			\includegraphics[valign=m,scale=0.75]{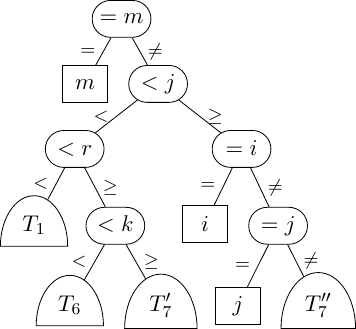}
			\caption{An example conversion of $T$ (on the left) into $T'$ (on the right)
			in Case~\ref{case 2: 3/7 bound}, where $k < j < i$. 
			 $T'_7$ and $T''_7$ are copies of $T_7$. 
			} 
			\label{fig: high lambda example}
		\end{figure}

		
		The idea is illustrated by the example in Figure~\ref{fig: high lambda example}.	
		In tree $T'$ we have two copies of $T_7$, denoted $T'_7$ and $T''_7$.
		(Because of this $T'$ is not irreducible. As explained in Section~\ref{sec: preliminaries}, $T'$ can be then
		converted into an irreducible tree by splicing out redundant branches.)
		These copies are needed because
		the range of $T_7$ is partitioned by the $\lessthantest{j}$ test into two subsets, with
		query keys smaller than $j$ following the $<$-branch and the other following the $\ge$-branch.
		We refer to this in text later as ``fracturing'' of $T_7$.
		However, since these subsets form a disjoint partition of the range of $T_7$, 
		the total contribution of $T'_7$ and $T''_{7}$ to the cost of $T'$ is the
		same as the contribution of $T_7$ to the cost of $T$. 
		More generally, such fracturing does not affect the cost as long as the fractured
		copies of a subtree are at the same depth as the original subtree.
		
		\smallskip
		
		Ultimately, using the case assumption that $w\parend{T_3} < \onethird w\parend{T_2}$, we want to show the following:
		
		\begin{claim}\label{cla: case 2 cost}
			There exists a tree $T'$ such that
			\[\cost(T')- \cost(T) \;\le\;  w\parend{T_1}  - w_m + 2w\parend{T_3}.\]
		\end{claim}
		
		In other words, we want to show that in the worst case scenario, we have a modified tree whose cost is at worst equivalent to moving key $m$ up by one, 
		$T_1$ down by one, and $T_3$ down by 2. 
		This suffices to prove Case~\ref{case 2: 3/7 bound} as then
		\begin{align*}
			\cost(T')-\cost(T) \;&\le\; w\parend{T_1}  - w_m + 2 w\parend{T_3}  \\
			&=\; w\parend{T} - 2w_m + w\parend{T_3} - w\parend{T_4} 
				 				&& w_m + w\parend{T_1} + w\parend{T_3} + w\parend{T_4} = w\parend{T} \\
			&\le\; w\parend{T} - 2w_m + \tfrac{1}{3}w\parend{T_2} - \tfrac{2}{3} w\parend{T_2} 
				 				&& w\parend{T_3} < \tfrac{1}{3}w\parend{T_2},  w\parend{T_4} > \tfrac{2}{3}w\parend{T_2}  \\
			&=\; w\parend{T}  - 2w_m - \tfrac{1}{3} w\parend{T_2} \\
			&\le\; w\parend{T} - 2w_m - \tfrac{1}{3} w_m 
				 					&& w\parend{T_2} \ge w_m  \\
			&\le\; 0
		\end{align*}

Before describing the construction of $T'$, we'll first establish  Claim~\ref{cla: bi's distinct} below.
		

\begin{claim}\label{cla: bi's distinct}
$r < b_2$ (so that $\angled{< r}$ and $\angled{< b_2}$ are distinct tests).
\end{claim}
		
Because keys $i$, $j$, and $k$ are in the ${\ge}$-branch of $\lessthantest{r}$, we have that $r \le b_1, b_2, b_3$. 
As any irreducible tree can perform at most two tests on the same key, if $b_i=r$ then $b_i$ is distinct from the other two keys. 
Then $b_i = b_1$ must hold to preserve order, and thus $b_2 > b_1 = r$. This proves Claim~\ref{cla: bi's distinct}.


\smallskip
The modified tree $T'$ then has the following form: $\equalstest{m}$ is at the root, $\lessthantest{b_2}$ is at the ${\ne}$-branch of $\equalstest{m}$, and $\lessthantest{r}$ is at the ${<}$-branch of $\lessthantest{b_2}$. Notably, $T_1$ will still be in the $<$-branch of $\lessthantest{r}$ in $T'$, 
which implies that $m$ moves up by 1 and $T_1$ moves down by 1, thus matching the $w\parend{T_1} - w_m$ terms in Claim~\ref{cla: case 2 cost}. 
The right branch of $\lessthantest{r}$ leads to $\unknowntest{b_1}$.
The rest of $T'$ will be designed so that all subtrees $T_3$, $T_5$, $T_6$ and $T_7$ will have roots
at depth at most $4$, so in particular $T_6$ and $T_7$ will never move down.

Then it suffices to show that the new depths of $T_3$ and $T_5$ imply a cost increase no greater than $2w\parend{T_3}$. More precisely, we will show that $T'$
has one of the following properties: Compared to $T$, in $T'$
\begin{enumerate}[label=(j\arabic*)] \setlength{\itemsep}{-0.05in}
	\item \label{case 2.2 condition 1} $T_3$ moves down by at most $2$  and $T_5$'s depth doesn't change, \emph{or}
	\item \label{case 2.2 condition 2} If $\unknowntest{j}$ is an equal-to test then $T_5$ and $T_3$ move down at most by $1$ each. 
	(This suffices because $w\parend{T_5} = w_j \le w\parend{T_3}$ if $\unknowntest{j}$ is an equal-to test,
	by applying Lemma~\ref{lemma: side-weight monoticity} to $T$.)
\end{enumerate}

To describe the rest of our modification, we break into two sub-cases, depending on whether $\unknowntest{b_2}$ is
and equal-to test or less-than test.


\begin{subcase}[$\unknowntest{b_2} = \lessthantest{b_2}$]\label{case: heaviest key case 2.1}
In this case, $\unknowntest{b_3}$ is at the $\ge$-branch of $\lessthantest{b_2}$ in $T'$. 
We do not introduce any new comparison tests, obtaining $T'$ shown in Figure~\ref{fig: high lambda case 2 fig c}. 
We now break into further cases based on which key $b_2$ is.
	
First, we observe that $b_2 \neq i$, as for $b_2 = i$ the structure of $T$ would imply that $j, k < i$, meaning that $i$ would be the largest key, instead
of the middle one. Thus, $b_2\in \braced{j,k}$. This gives us two sub-cases.

\begin{description} \setlength{\itemsep}{0.02in}
\item{\textbf{Case 2.1.1:}} $b_2= j$. Then we have $k < j$ by the structure of $T$, implying $k =b_1$, and $i \ge j$ since $i$ must now be $b_3$. 
Then, in $T'$,
$\unknowntest{b_1}$ has branches $T_6$ and $T_7$, while $\unknowntest{b_3}$ has branches $T_3$ ($\EqOrGe$-branch) and $T_5$ ($\NeOrLe$-branch). In which case, only $T_3$ moves down by 1, satisfying \ref{case 2.2 condition 1}.
\item{\textbf{Case 2.1.2:}} $b_2= k$. Then the structure of $T$ implies that $k\neq\braced{i,j}$, so $b_1 < k$, which in turn implies that $\unknowntest{b_1}$ is an equal-to test.
We now have two further sub-cases.
If $(i,j) = (b_1,b_3)$ then, in $T'$, $T_5$ is in the $\EqOrGe$-branch of $\unknowntest{b_3}$ and leaf $T_3  = i$ is
in the $\EqOrGe$-branch of $\unknowntest{b_1}$, implying $T_5$'s depth doesn't change and $T_3$ moves down by 2, satisfying \ref{case 2.2 condition 1}. 
If $j = b_1$, then leaf $T_5 = j$ is below $\unknowntest{b_1}$ and $T_3$ is below $\unknowntest{b_3}$, both moving down by 1, satisfying \ref{case 2.2 condition 2}.
\end{description}
\end{subcase}


		\begin{figure}[t]
			\centering
			\begin{subfigure}[b]{0.48\textwidth}
				\centering
				\includegraphics[scale=0.75]{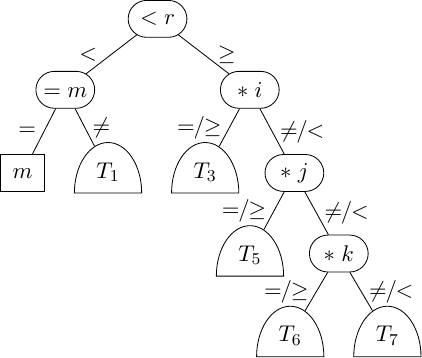}
				\caption{Original tree $T$}\label{fig: high lambda case 2 fig a}
			\end{subfigure}~
			\begin{subfigure}[b]{0.48\textwidth}
				\centering
				\includegraphics[scale=0.75]{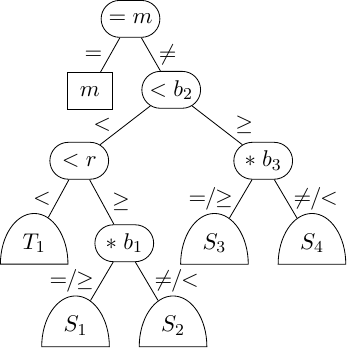}
				\caption{Modified tree $T'$}\label{fig: high lambda case 2 fig c}
			\end{subfigure}
			\caption{Original and modified tree for Case~\ref{case: heaviest key case 2.1}. $S_1 , S_2 , S_3 , S_4$ is simply some permutation of $T_3, T_5, T_6, T_7$.} \label{fig: high lambda case 2.1}
		\end{figure}
		
		\begin{figure}[t]
				\centering
				\begin{subfigure}[b]{0.32\textwidth}
					\centering
					\includegraphics[scale=.75]{highlambda-case2-figa.pdf}
					\caption{Original tree $T$}
				\end{subfigure}~
				\begin{subfigure}[b]{0.32\textwidth}
					\centering
					\includegraphics[scale=.75]{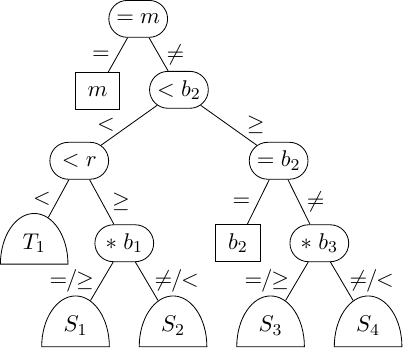}
					\caption{Modified tree $T'$, $\equalstest{b_2}$ goes first}\label{fig: high lambda case 2 fig b}
				\end{subfigure}
				\begin{subfigure}[b]{0.32\textwidth}
					\centering
					\includegraphics[scale=.75]{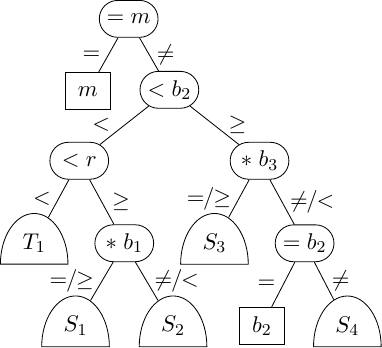}
					\caption{Modified tree $T'$, $\unknowntest{b_3}$ goes first}\label{fig: high lambda case 2 fig d}
				\end{subfigure}
				\caption{Original and modified tree for Case~\ref{case: heaviest key case 2.2}. 
				In this case, $T_5 = b_2$, and  $S_1, S_2, S_3, S_4$ is some permutation of $T_3, T_6, T_7, T''$, 
				where $T''$ is a copy of $T_6$ or $T_7$. } \label{fig: high lambda case 2.2}
			\end{figure}


\begin{subcase}[$\unknowntest{b_2} = \equalstest{b_2}$ and both tests $\unknowntest{b_1}$, $\unknowntest{b_3}$ are different from $\lessthantest{b_2}$]
\label{case: heaviest key case 2.2}
In this case, $\lessthantest{b_2}$ (the $\neq$-child of $\equalstest{m}$ in $T'$) is a newly introduced test.
In $T'$, we will have $\equalstest{b_2}$ and $\unknowntest{b_3}$ in the $\ge$-branch of $\lessthantest{b_2}$. 
The order in which we perform  $\equalstest{b_2}$ and $\unknowntest{b_3}$ will be determined later.

We will say $\equalstest{b_2}$ is ``performed first'' if it is the root of the $\ge$-branch of $\lessthantest{b_2}$, in which
case $\unknowntest{b_3}$ is rooted at the $\neq$-branch of $\equalstest{b_2}$, or ``performed second''.
Likewise if $\unknowntest{b_3}$ is performed first, then $\equalstest{b_2}$ is performed second, rooted at the $\NeOrLe$-branch of $\unknowntest{b_3}$.
Figure~\ref{fig: high lambda case 2.2} illustrates these two possible configurations.
In most cases, $\equalstest{b_2}$ and $\unknowntest{b_3}$ are performed in the same order as in the original tree $T$ 
(i.e. $\unknowntest{i}$ comes before $\unknowntest{j}$, which comes before $\unknowntest{k}$), 
though one case (Case~2.2.2) requires going out-of-order, implying a rotation.
	
Because $\lessthantest{b_2}$ is a new comparison, one of the sub-trees $T_3$, $T_5$, $T_6$, $T_7$ in $T$ may fracture, meaning that some of
its keys may satisfy this comparison and other may not. We will in fact show that only $T_6$ or $T_7$ can fracture, 
but their depths do not increase. So, as explained earlier, this fracturing will not increase cost.
(As also explained before, the redundancies created by this fracturing, and other that can occur as a result of the conversion,
can be eliminated by post-processing $T'$ that iteratively splices out redundant branches.)
We again break into further cases based on which key $b_2$ is.

\begin{description} \setlength{\itemsep}{0.02in}
\item{\textbf{Case 2.2.1:}} $b_2 = i$ and $\unknowntest{j}$ is an equal-to test.
Then $T_3 = i$ and $T_5 = j$ are both leaves and can't fracture. 
Perform $\equalstest{b_2}$ first and $\unknowntest{b_3}$ second (since $i=b_2$, this follows order of comparisons in the original tree).
Then $T_3$ and $T_5$ both move down by 1, satisfying~\ref{case 2.2 condition 2}, 
with $T_3 = i = b_2$ in the $=$-branch of $\equalstest{b_2}$ and $T_5 = j$ in the $=$-branch of either $\unknowntest{b_1}$ or $\unknowntest{b_3}$ (depending on whether $j$ is $b_1$ or $b_3$).
\item{\textbf{Case 2.2.2:}} $b_2 = i$ and $\unknowntest{j}$ is a less-than test. 
Then $k < j$, so $j=b_3$ and $k=b_1$. 
By the case assumption, we have that $j > i$. Then in the $\ge$-branch of $\lessthantest{b_2}$, we perform $\lessthantest{j}$ \emph{before} $\equalstest{i}$ (going out-of-order compared to the original tree), which will be in the $<$-branch of $\lessthantest{j}$. Then $T_3 = i$ will be below $\equalstest{i}$ moving down twice and $T_5$ will be in the $\ge$-branch of $\lessthantest{j}$ staying at the same depth, satisfying \ref{case 2.2 condition 1}.
\item{\textbf{Case 2.2.3:}} $b_2 = j$.
Then we may simply perform $\equalstest{b_2}$ and $\unknowntest{b_3}$ in the same order as in the original tree. 
If $\unknowntest{i}$ is an equal-to test, then $T_3$ and $T_5$ are both leaves, and either both will go down by 1 (if $i=b_3$ and $k=b_1$), satisfying~\ref{case 2.2 condition 2},
or only $T_3$ goes down by 2 (if $i=b_1$ and $k=b_3$), satisfying~\ref{case 2.2 condition 1}. 
If $\unknowntest{i}$ is a less-than test, then $j, k < i$, so $k = b_1$ and $i= b_3$. 
$\lessthantest{i}$ will be performed first in the $\ge$-branch of $\lessthantest{b_2}$ and $T_3$ will be in the $\ge$-branch of $\lessthantest{i}$, 
implying $T_3$ and $T_5$ both go down by 1, satisfying~\ref{case 2.2 condition 2}.
\item{\textbf{Case 2.2.4:}} $b_2 = k$. 
The analysis is similar to Case 2.1.2. By the structure of $T$, $k \neq \braced{i, j}$, so $b_1 < k$, implying $\unknowntest{b_1}$ is an equal-to test. 
Then perform $\equalstest{b_2}$ and $\unknowntest{b_3}$ in the same order as the original tree (that is, $\equalstest{k}$ is always performed second). 
Thus $\unknowntest{b_1}$ is in the $\ge$-branch of $\lessthantest{r}$ and $\unknowntest{b_3}$ is in the $\ge$-branch of $\lessthantest{b_2}$. 
If $(i, j) = (b_1 , b_3)$, then $T_3$ (which is a leaf $i$) is at the $\EqOrGe$-branch of $\unknowntest{b_1}$ and $T_5$ is at the $\EqOrGe$-branch of $\unknowntest{b_3}$, 
implying $T_5$'s depth stays the same and $T_3$ moves down by 2, satisfying \ref{case 2.2 condition 1}. If $(i, j) = (b_3, b_1)$, 
then leaf $T_5 = j$ is below $\unknowntest{b_1}$ and $T_3$ is below $\unknowntest{b_3}$, implying both trees move down by 1, satisfying~\ref{case 2.2 condition 2}.
\end{description}

\qedhere
\end{subcase}

	\end{case}